\pdfoutput=1

\newcommand{\beq}{\begin{equation}}
\newcommand{\eeq}{\end{equation}}
\newcommand{\beqnl}{\begin{eqnarray}}
\newcommand{\beqna}{\begin{eqnarray*}}
\newcommand{\eeqna}{\end{eqnarray*}}
\newcommand{\eeqnl}{\end{eqnarray}}

\documentclass[]{iopart} 	
\usepackage{amsfonts}
\usepackage{setstack}
\usepackage{amsthm}
\usepackage{amsbsy}
\usepackage{amssymb}
\usepackage{graphicx}
\usepackage{subfigure}
\begin{document}

\date{\today}

\title[]{Spacetime geometries and light trapping in travelling refractive index perturbations}

\author{S.L.~Cacciatori$^{1,2}$, F.~Belgiorno$^{3}$, V.~Gorini$^{1,2}$, G.~Ortenzi$^{4}$, L.~Rizzi$^{1}$, V.G.~Sala$^{1}$,   D.~Faccio$^{5}$}
\address{$^1$Department of Physics and Mathematics, Universit\`a dell'Insubria, Via Valleggio 11, IT-22100 Como, Italy\\
$^2$ INFN sezione di Milano, via Celoria 16, IT-20133 Milano, Italy\\
$^3$Dipartimento di Fisica, Universit\`a di Milano, Via Celoria 16, IT-20133 Milano, Italy\\
$^4$Dipartimento di Matematica e Applicazioni, Universit\`a degli Studi di Milano, Bicocca, IT-20125 Milano, Italy\\ $^5$CNISM and Department of Physics and Mathematics, Universit\`a dell'Insubria, Via Valleggio 11, IT-22100 Como, Italy\\ 
E-mail: \url{sergio.cacciatori@uninsubria.it} \\}

\begin{abstract}
In the framework of transformation optics, we show that the propagation of a locally superluminal
refractive index perturbation (RIP) in a Kerr medium can be described,
in the eikonal approximation, by means of a stationary metric which we prove to be of Gordon type.
Under suitable hypotheses on the RIP, we obtain a stationary but not static metric, which is
characterized by an ergosphere and by a peculiar behaviour of the geodesics, which are studied
numerically, also accounting for material dispersion.
Finally the equation to be satisfied by an event horizon is also displayed and briefly discussed.
\end{abstract}

\maketitle

\section{Introduction}
\label{intro}

Effective geometries for light, first introduced by Gordon \cite{gordon} and extended to nonlinear electrodynamics, are able to provide analogue black hole metrics and also the possibility to perform experiments involving Hawking analogue radiation
\cite{Leonhardt:1999fe,Leonhardt:2000fd,Schutzhold:2001fw,Brevik:2001nf,DeLorenci:2001ch,DeLorenci:2001gf, Novello:2001gk,Marklund:2001dq,DeLorenci:2002ws,Novello:2002ed,Novello:2003je,Unruh:2003ss, Schutzhold:2004tv}.
Recently Philbin et al. proposed an optical analogue in which a soliton with intensity $I$, propagating in an optical fibre, generates through the nonlinear Kerr effect a refractive index perturbation (RIP), $\delta n=n_2 I$, where $n_2$ is the Kerr index \cite{philbin}. The same mechanism has also been generalized to a full 4D geometry by Faccio et al. \cite{faccio}. The RIP modifies the spacetime geometry as seen by co-propagating light rays and, similarly to the acoustic analogy, if the RIP is locally superluminal, i.e. if it travels faster than the phase velocity of light in the medium, a trapping horizon is formed and Hawking radiation is to be expected. A full analysis for the case of a static black hole metric has been recently discussed in \cite{rip-static}.
Our aim here is to tackle the following aspects:
(i) provide a general derivation of the metric in the eikonal approximation from
relativistic nonlinear electrodynamics in a medium and (ii) perform a first analysis for a generic RIP
which gives rise to a stationary (but in general non static) metric. The
metric we obtain allows to determine in a straightforward way the limits under which
an ergoregion occurs.
Then a study of the behaviour of null geodesics follows, with the aim of identifying
conditions such that trapping of light occurs inside the dielectric perturbation.
Finally, the equation to be satisfied by an event horizon is provided. 

\section{Effective geometry}

In this section, we introduce the specific analogue model that we have analysed. As we need to work both in the laboratory frame as well as in an inertial reference frame which is moving at relativistic speed with respect to the original frame, it is convenient to adopt a covariant formalism. Then, consider a reference frame in which the dielectric medium is moving with four-velocity $u^\mu$. Moreover, consider a medium for which the permittivity and the permeability have respectively the form
\beqnl
&&\epsilon^{\alpha\beta} = \varepsilon(E)(\eta^{\alpha\beta} - u^\alpha u^\beta) , \label{epsilon}\\
&&\mu^{\alpha\beta} = \mu_0\mu_r (\eta^{\alpha\beta} - u^\alpha u^\beta) \label{mu},
\eeqnl
where $\eta_{\mu\nu}$ is the Minkowski metric tensor.
That is, the permittivity is a function of the invariant electric field amplitude $E := \sqrt{ - E^\mu E_\mu }$ alone. In the laboratory frame, this reduces to the assumption that a polarisation is induced in the medium only when it is exposed to an electric field, and that the nonlinear effects depend only on the intensity of the latter. The permeability is assumed to be constant and equal to the product $\mu$ of the vacuum permeability $\mu_0$ and the relative permeability $\mu_r$ of the medium. These conditions indeed stand true for diverse nonlinear media. For example, for the case of an isotropic $\chi^{(3)}$ medium, in the laboratory frame
\begin{equation}
\varepsilon = \varepsilon_0\varepsilon_r = \varepsilon_0(1 + \chi^{(1)} + \chi^{(3)}E^2),
\end{equation}
where $\varepsilon_0$ is the vacuum permittivity.

Nonlinear effects are usually very small and a very strong field is needed to exploit the nonlinearity of the medium itself. Thus, we assume that there exists a strong, background field $E$, which fixes the properties of the dielectric. Note that these assumptions are rather general. In fact, Eqs.~(\ref{epsilon}) and~(\ref{mu}) hold for any nonlinear, homogeneous dielectric, no matter how the strong, background pulse is generated \cite{DeLorenci:2001ch}. We are interested in the small and rapidly oscillating fluctuations of the electromagnetic field on top of such a background\footnote{For reasons connected with the actual experimental verification of this theoretical framework, we may think of this strong background field as a spatially localized (albeit rather extended) pulse, which is travelling through the medium.}.

According to the results of Refs.~\cite{DeLorenci:2001ch,Novello:2002ed,Novello:2003je}, these small fluctuations, ``feel'' an effective curved spacetime. The result is what a relativist would call a ``bi-metric" theory. In other words, there exist two classes of rays, namely ordinary and extra-ordinary rays, which correspond to two different polarizations of the background field. This phenomenon is analogous to birefringence in uni-axial crystals, where the ordinary and extraordinary rays each obey distinct quadratic dispersion relations. In the analogue gravity terminology, each ray ``feels" a different geometry, or \emph{analogue metric}: the Gordon's metric
\begin{equation}\label{g}
g_{\mu\nu}^{(+)} = \eta_{\mu\nu} - u_\mu u_\nu\left[1-\frac{1}{n^2}\right]
\end{equation}
for the ordinary rays, and the metric
\begin{equation}\label{eg}
g_{\mu\nu}^{(-)} = \eta_{\mu\nu} - u_\mu u_\nu\left[1-\frac{1}{n^2(1+\xi)}\right] + \frac{\xi}{\xi +1}l_\mu l_\nu
\end{equation}
for the extra-ordinary rays. In (\ref{g})-(\ref{eg}), we have introduced the ``nonlinear parameter'' $\xi:=\frac{E}{\varepsilon}\frac{d \varepsilon}{dE}$, the background field polarization four-vector $l_\mu :=\frac{E_\mu}{E}$, and the refractive index $n:=\sqrt{\varepsilon_r\mu_r}$.
In cartesian coordinates $(t,x,y,z)$, $\eta_{\mu\nu} = \text{diag}(c^2, -1,-1,-1)$.

In \ref{appA} we show that, under reasonable approximations, the extra-ordinary metric is equivalent to the Gordon one, albeit with a different effective Kerr perturbation of the refractive index, i.e. $n=n_0+3dn$ instead of
$n=n_0+dn$. Thus, we restrict our attention to the Gordon metric.

In the following we study the lightlike geodesics of the analogue metrics. It is important to note that these geodesics describe the spacetime trajectories of monochromatic waves. In the present paper we often refer to these trajectories with the term ``light rays''. However these should not be confused with the usual light rays referred to in geometrical optics which, in the Newtonian picture, describe the trajectories of wave packets (or ``photons''). For this reason, in the forthcoming sections, we analyse the monochromatic components (congruences of lightlike geodesics) of the electromagnetic field and the analogue geodesic motion of the related constant phase surfaces, which are precisely the objects involved in the study of the analogue Hawking radiation\cite{belg-black, PhysRevA.81.063835}. By contrast we do not consider the notion of ``photon'' (or wave packet) here.

\section{The metric in the RIP frame}

Consider the case of a pulse moving with uniform velocity $v$ along the positive $x$ axis. According to the discussion above, the pulse excites
a localized RIP which also moves accordingly. The latter can then be analytically described as a function of $x-vt$ and, assuming cylindrical symmetry, of the transversal coordinate $\rho=\sqrt {y^2+z^2}$. Thus, in the laboratory frame, $n=n_0+\delta n (x-vt,\rho)$. In this situation, it is convenient to express the Gordon metric in a reference frame which is co-moving with the RIP itself:
\beqnl
ds^2 &=& c^2 \gamma^2 \frac{1}{n^2} \left(1+\frac{n v}{c}\right)\left(1-\frac{n v}{c}\right) dt'^2 +
2 \gamma^2 \frac{v}{n^2} \left(1-n^2\right) dt' dx' \cr
&&- \gamma^2 \left(1+\frac{v}{n c}\right)\left(1-\frac{v}{n c}\right) dx'^2-d\rho^2 -\rho^2 d\phi^2,\label{metrica-primata}
\eeqnl
where the primed coordinates are relative to the co-moving frame, and
\beq
n : = n_0+\delta n (x',\rho).
\eeq
Here, $\delta n$ is a $C^{\infty}$ function, rapidly decaying at infinity and with a single maximum of height $\eta$, describing the RIP.
It is then evident that both $\partial_{t'}$ and $\partial_{\phi}$ are Killing vectors for
the given class of metrics. Any specific choice for the function $\delta n$ gives rise to a specific
metric in the aforementioned class.
We point out that an isotropic refractive index
in the laboratory frame corresponds to an anisotropic refractive index in the pulse frame due to
length contraction associated with a boost in the $x$ direction.
If the refractive index depends explicitly on $\rho$, the metric is stationary but not static:
the integrability conditions stated in Frobenius theorem \cite{wald} are not satisfied. In particular, if
$\xi_{\mu}=(g_{00},g_{01},0,0)$ are the covariant components of the timelike Killing vector $\partial_{t'}$,
the integrability conditions $\xi_{[\lambda} \nabla_{\mu} \xi_{\nu]} =0$, for $\lambda=0,\mu=1$
give $\xi_{0} (\nabla_{1} \xi_{\nu}-\nabla_{\nu} \xi_1) + \xi_{1} (\nabla_{\nu} \xi_0 -\nabla_{0} \xi_{\nu})=0$,
i.e., for $\nu=\rho$, $g_{00} (-\partial_{\rho} g_{01}) +  g_{01} \partial_{\rho} g_{00}=0$, which is easily seen to be not satisfied. The static case is obtained if $\delta n$ does not depend on $\rho$, and it has been extensively
studied in \cite{rip-static}. Here, we consider the more general stationary, non-static case.\footnote{A simple, non-technical definition of static and stationary spacetimes can be given as follows. A spacetime is stationary if it is possible to find a reference frame in which the metric coefficients do not depend on time. This means that time translation is a symmetry of the spacetime. However, such a spacetime might not be invariant under time-reversal. In the case of time-reversal invariance, the spacetime is not only stationary, but also static. An example of static spacetime is the Schwarzschild solution, describing the static empty spacetime outside a non-rotating black hole. If the black hole has a non-vanishing angular momentum, the spacetime is still invariant under time translations, however, time reversal reverses the sign of the angular momentum, turning a clockwise spinning black hole to an anticlockwise one. Then, the spacetime around a spinning black hole is stationary, but not static.}
We first point out that for our class of metrics an ergosurface is allowed, given by the locus of points satisfying the equation
\beq\label{ergo}
g_{00} = 0 \Leftrightarrow 1-\frac{n v}{c}=0.
\eeq
This equation admits solutions only for the range of RIP velocities $v$ given by
\beq
\frac{1}{n_0+\eta} \leq \frac{v}{c} < \frac{1}{n_0}.
\label{cond-ergo-eta}
\eeq
At the ergosurface, which is the boundary of the ergoregion,
the Killing vector $\partial_{t'}$ vanishes and upon entering the ergoregion it changes sign.
Here, in contrast with the static case,
the ergosurface does not coincide with the event horizon. Indeed,
the given hypersurface is not a lightlike one, as can be easily verified by looking at
the behaviour of its normal vector. However, for $\rho=0$ it actually becomes lightlike, i.e. it meets
the event horizon in the antipodal points which solve
\beq
\delta n(x',0)=\frac{c-n_0 v}{v}.
\eeq
For stationary nonstatic situations it is more difficult to identify a nontrivial event horizon in
effective geometries (see e.g.  Visser's analysis for the case of acoustic black holes \cite{visser-acou}).
Before discussing the question of the existence of an event horizon,
we investigate whether the path of light rays, that is the null geodesic curves of metric (\ref{metrica-primata}),
do suggest the existence of such a horizon.

\section{Geodesics and trapping of light}

In the laboratory frame the metric (\ref{metrica-primata}) is
\beq
ds^2  = \frac{c^2}{n^2 (x-vt,y,z)}dt^2 -dx^2 -dy^2 -dz^2 .\label{laboratory}
\eeq
We model the RIP with a gaussian shaped function which propagates along the $x$ axis with velocity $v$. We also suppose that, during the propagation, it keeps its shape both in space and in time.
Assuming cylindrical symmetry, we need to account only for one of the two transverse coordinates, y, say. Then, we write the refractive index
in the form
\begin{eqnarray}
n = n_0+\delta n, \qquad
\delta n=\eta e^{\frac{-y^{2}-(x-vt)^{2}}{\sigma^2}}.\label{gauss}
\end{eqnarray}
Thus, the geodesic equations for null rays take the explicit form
\begin{eqnarray}
&\ddot x&=-\frac 2{\sigma^2}\frac{\delta n}{n^3}(x-vt) c^2\dot t^2 , \label{eq1}\\
&\ddot y&=-\frac 2{\sigma^2} \frac{\delta n}{n^3}y c^2 \dot t^2 , \label{eq2}\\
&\ddot t&=\frac 2{\sigma^2} \left[ \frac{\delta n}{n^2}v(x-vt)\dot t^2- y \dot y \dot t- (x-vt)\dot x \dot t\right], \label{eq3}
\end{eqnarray}
where the dot denotes derivation w.r.t. an affine parameter $\lambda$.
A trivial solution is $\frac{dx}{dt}=c/n$, $y(t)=0$.
However, in the general case a numerical analysis is required.
In the following subsection we solve the equations above using the Matlab's ode45 function.
This analysis will provide an insight in the underlying physics.
\subsection{Numerical solutions and trapping of light}
\begin{figure}[ht]
\centering
\includegraphics[width=0.8\textwidth]{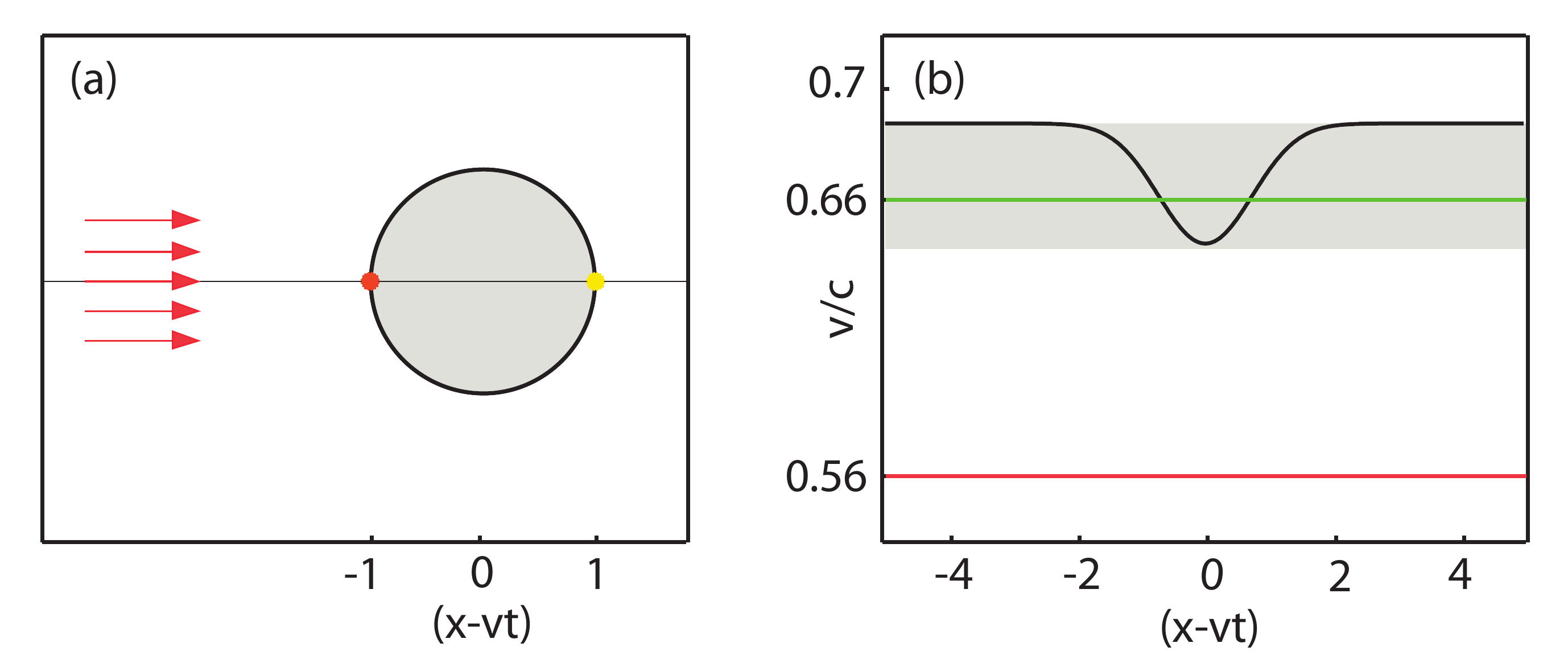}
\caption{(a) Sketch showing the initial conditions of the simulation. (b) Light ray phase velocity ($c/n$). The shaded region highlights the range of velocities of the RIP for which the rays are trapped. The two straight lines indicate the values of $v/c$ used in the simulations.}\label{velo}
\end{figure}

Along a geodesic we have
$(c\dot t,\dot x, \dot y,\dot z )=(k^0,k^1,k^2,k^3)=(n^2\omega/c,k_x,k_y,k_z),$
and taking the first derivative (with respect to the affine parameter $\lambda$) of the relation
$c\dot t=n^2\omega/c$,
we find an equation for $\dot \omega$ which determines the evolution of the frequency of the ray along a geodesic.
At this point a comment is in order: here we have defined $\omega$ as the zero-th covariant component of the four-momentum $k^\mu$.
It is easy to see that this corresponds to measuring frequencies as the inverse of the lapse of coordinate time between two successive
crests (this time is the Minkowskian time measured by the laboratory observer). This is of course different from the analogue proper time
$d\tau=ds/c$ of the analogue metric (\ref{laboratory}), which has no physical significance: experimental
measurements refer to the true Minkowski metric and not to the analogue curved metric.

Then, for the evolution of the frequency $\omega$ along the path of the ray, we obtain
\begin{equation}
\dot \omega=\frac{\ddot t-2\omega n \dot \delta n}{n^2}. \label{omega}
\end{equation}
For the sake of definiteness assume the propagation to be in fused silica and the rays to have initial wavelength $\lambda_{in}=527$ nm.
We assume the RIP to have $\sigma=1$ m and amplitude $\eta=0.1$, which is a quite unrealistic situation, but simple to deal with. Indeed, the aim of the simulation is merely to study the qualitative behaviour of the geodesics. We have verified that this behaviour does not change even for much smaller and realistic values of $\sigma\sim 1-100$ $\mu$m.

The other parameters of the simulation are the velocity $v$ of the RIP and the initial conditions of the rays, i.e. the array $\{t_{in},x_{in},y_{in},\dot t_{in},\dot x_{in},\dot y_{in}\}$. In the simulation, the rays start away from the RIP (which is right moving)
and their initial velocity is directed along the $x$ axis, as shown in the sketch of Fig.~\ref{velo}(a). More precisely, the initial conditions are $\{0,x_{in},y_{in},\dot t_{in},\dot x_{in},0\}$, with the additional constraints $\dot x_{in} / \dot t_{in}=c/n_{0}$ and $c\dot t_{in}=n_{0}^2\omega_{in}$.

The results of the simulation show that the behaviour of null geodesics is strongly dependent on the velocity $v$ of the RIP. Namely, there exists a range of velocities for which the rays are trapped \footnote{This is true also when taking into account the material dispersion, which is neglected in this subsection.}, and these coincide with those predicted by Eq.~(\ref{cond-ergo-eta}) (shaded region in Fig.~\ref{velo}(b)).
On the other hand, for lower values of $v$, the rays which enter the RIP are bent and eventually escape.
In Fig.~\ref{velo}(b), the black curve represents the velocity of an axial ray, i.e. of a ray propagating along the $x$ axis, versus $x-vt$. In the same figure, we represent with horizontal lines two choices for the RIP velocity, representing two possible situations: a faster (green line) one and a
slower (red line) one. In the first case (faster RIP), the green line intersects the black curve.
As the ray approaches the point corresponding to the first intersection it slows down due to the increasing refractive index, until it acquires the same velocity of the RIP. Asymptotically, the ray is ``trapped'', and moves together with the RIP. In the other case (slower RIP), the ray velocity is always larger then the RIP's one (red line of Fig.~\ref{velo}(b)). Then, the ray can cross the RIP and escape. In this ``subluminal'' case, no trapping occurs.

\begin{figure}[ht]
\centering
\includegraphics[width=0.8\textwidth]{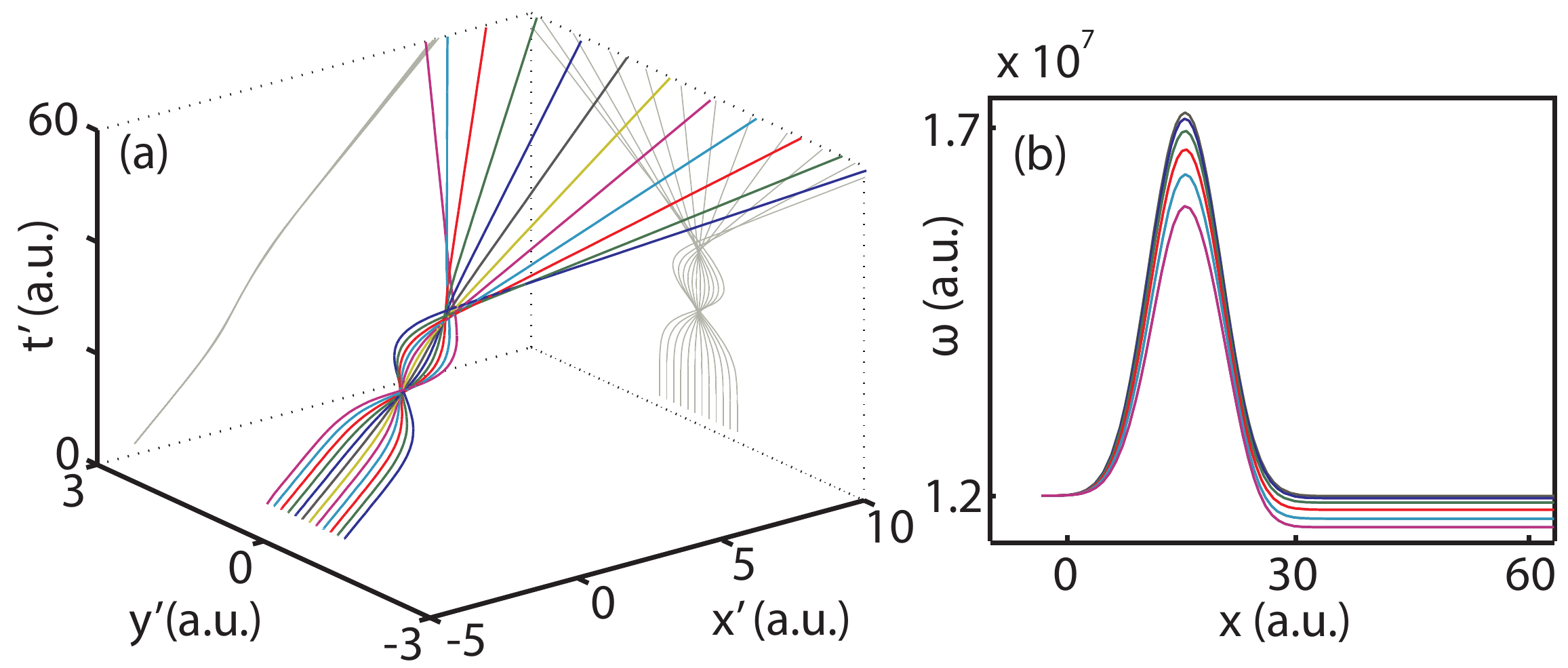}
\caption{Numerical results in the co-moving reference frame (a) and frequency evolution in the laboratory reference frame (b) for the subluminal case. }\label{0p56}
\end{figure}
\noindent \emph{Subluminal RIP}
The RIP propagates at $v=0.56c$ (red line in Fig.~\ref{velo}(b)), and it is always slower than the light rays. In Fig.~\ref{0p56}(a) we show the numerical results of the simulation in the co-moving reference frame.
The rays start on the left of the RIP. They propagate along straight lines with velocity $c/n_0$ until they reach the RIP. Then, inside the RIP, they slow down, bend and finally escape from the RIP, propagating along straight lines in different directions, depending on their impact parameter $y_{in}$. The projection of the trajectory on the $(t',x')$ plane, which is also shown Fig.~\ref{0p56}(a), evidently shows a change of slope
corresponding to the trajectory being inside the RIP. This means that the rays, while crossing the RIP, slow down.
Fig.~\ref{0p56}(b) represents the evolution of $\omega$ along the $x$ direction in the laboratory reference frame. We observe that, as expected, the frequency varies consistently when the rays reach the RIP. The rays go through a blueshift inside the RIP and then they emerge with a Doppler redshift, due to the fact that the rays are bent, and thus emerge at different angles. This behavior is described in more detail in \cite{faccio}.
\

\noindent \emph{Trapping RIP:}
\begin{figure}[ht]
\centering
\includegraphics[width=0.8\textwidth]{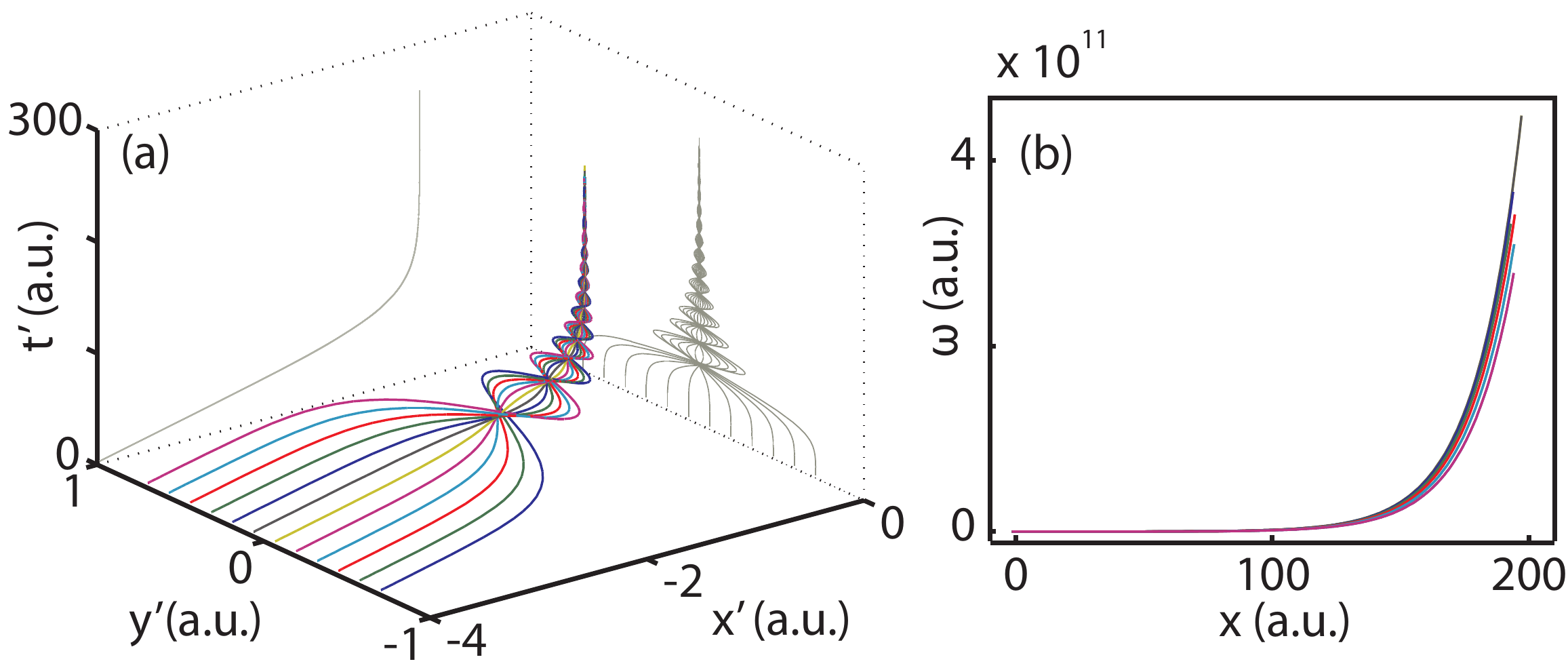}
\caption{Numerical results in the co-moving reference frame (a) and frequency evolution in the laboratory reference frame (b) for the trapping  case.}\label{0p66}
\end{figure}
The RIP propagates at $v=0.66c$ (green line in Fig.~\ref{velo}(b)). In Fig.~\ref{0p66}(a) we show the numerical results in the co-moving reference frame. The rays propagate initially along straight lines with velocity $c/n_0$ (for clarity, we recall that rays propagate along geodesics with the \emph{phase} velocity of the corresponding electromagnetic field). When they reach the RIP, they are bent and at the same time slow down. As shown in Fig.~\ref{velo}(b), at a certain point, away from the center of the RIP, the rays will have exactly the same velocity as the RIP,
and can no longer escape from it. In the co-moving reference frame they continuously slow down and at the same time they oscillate in the transverse direction with decreasing amplitude oscillations, until, asymptotically, all of them stop at the same point. This point, represented in red in Fig.~\ref{velo}(a), is the white hole horizon $x_-$ (see next section). If we look at the slope of the projection $t'=t'(x')$ we see that it diverges, which means that the rays velocity tends to zero corresponding to this point. Clearly, in the laboratory reference frame the rays do not stop but they move together with the RIP. As shown in Fig.~\ref{0p66}(b), as the rays reach the RIP, their frequency starts growing indefinitely, so we have an infinite blueshift.

\subsection{Effects of dispersion}
In the above considerations we did not account for the material dispersion, that is we assumed $n(\mathbf{r},t)$ independent of $\omega$. However, the refractive index of a medium is in general a complicated function of the frequency $\omega$ of the incident wave, that is $n=n(\mathbf{r},t,\omega)$. We restrict our attention to the case of ``normal dispersion'', that is we assume that the refraction index grows with frequency. In this case, the full dependence of the refractive index from $\omega$ is phenomenologically described by the Sellmeier equation. A standard way of treating dispersion is to expand $n(\omega)$ in power series of $\omega$ (about an initial frequency $\omega_0$), or equivalently, by the relation $c k=\omega n$, to expand $k$ in power series of $\omega$ as follows:
\begin{eqnarray}\label{n-expans}
n(\omega) = \frac{c}{\omega} \left( k(\omega_0)+ \sum_{n=1}^\infty \frac 1{n!} \frac{d^{n}k}{d\omega^{n}}\Big{|}_{\omega_0} \Delta \omega^n  \right).
\end{eqnarray}
The material dispersion also affects the metric, which therefore depends on the frequency:
\begin{equation} \label{metrica-disp}
ds^2 = \frac{c^2}{n^2(\mathbf{r},t,\omega)} dt^2 - d\mathbf{r}^2.
\end{equation}
Therefore, different rays propagating in the medium ``feel'' different metrics, depending on their frequency. Moreover, even a single ray will see an evolving metric while travelling, since its frequency (and hence the effective metric) changes during the propagation.
The geodesic equations (\ref{eq1}-\ref{eq3}) keep the same form as in the non-dispersive case, while Equation (\ref{omega}) now takes the form
\begin{equation}
\dot \omega=\frac{c^2\ddot t-2\omega n \dot \delta n}{n(n+2\omega dn_0/d\omega)}.
\end{equation}
We highlight that, contrary to the non-dispersive case, the evolution of the geodesics depends on the change in frequency of the ray, owing to the explicit dependence on $\omega$ in the geodesic equations.
\begin{figure}[ht]
\centering
\includegraphics[width=0.8\textwidth]{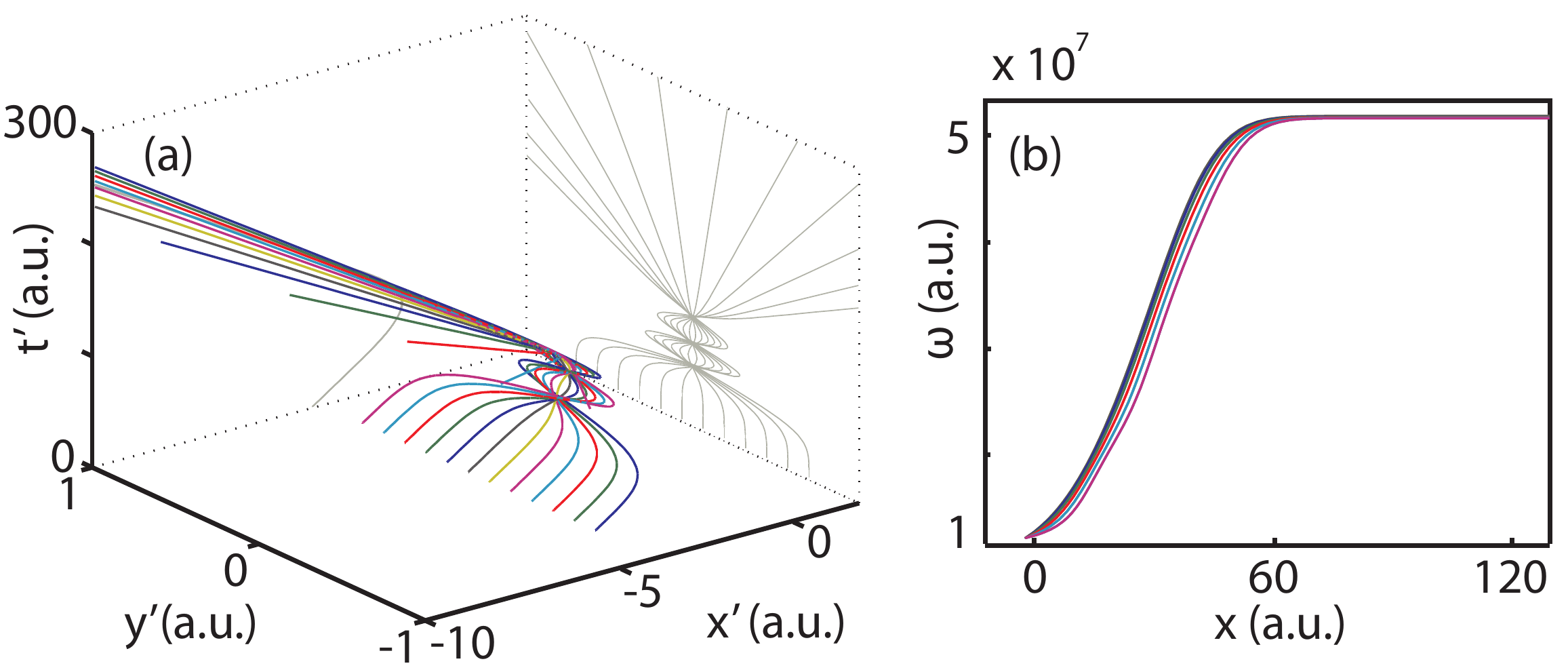}
\caption{Numerical results in the co-moving reference frame (a) and frequency evolution in the laboratory reference frame (b) for the trapping  case when we account for dispersion.}\label{v0p66disp}
\end{figure}
In the case of a subluminal RIP,
dispersion does not modify the qualitative behaviour of the light ray, which remains very similar to the dispersion-less case shown in Fig.~\ref{0p56}. On the other hand, for a trapping RIP, the presence of dispersion modifies substantially the geodesics.
The results of the numerical simulation are shown in Fig.~\ref{v0p66disp}. The trapping feature of the horizon is still evident.
However, taking dispersion into account, the rays are no longer indefinitely trapped inside the RIP, and eventually escape. This behaviour is clearly
depicted in Fig.~\ref{v0p66disp}(a), where the results are shown in the co-moving reference frame. Note that, at the beginning the (phase)  velocity of the rays is positive, which means that the rays are faster than the RIP, but then it becomes negative, implying that the rays have become slower than the RIP. Let us clarify this aspect. The phase velocity of a ray is defined as $v_{r}=\omega / k$. Under conditions of normal dispersion, i.e.
$\frac {d^2 k}{d\omega^2}>0$, as in our case, $k$ grows faster than linearly with $\omega$ (while without dispersion it would have a linear growth). When the rays approach the RIP, their frequency grows and, at the same time, their phase velocity decreases. Thus, the rays will begin to lag behind with respect to the RIP. When they are out of the RIP they will keep their final frequency. In Fig.~\ref{v0p66disp}(b) we show that the frequency grows only by a finite amount. Therefore, the rays leave the RIP with a finite (albeit very large) blueshift.\\

\noindent \emph{Trapping time   }
\begin{figure}[ht]
\centering
\includegraphics[width=\textwidth]{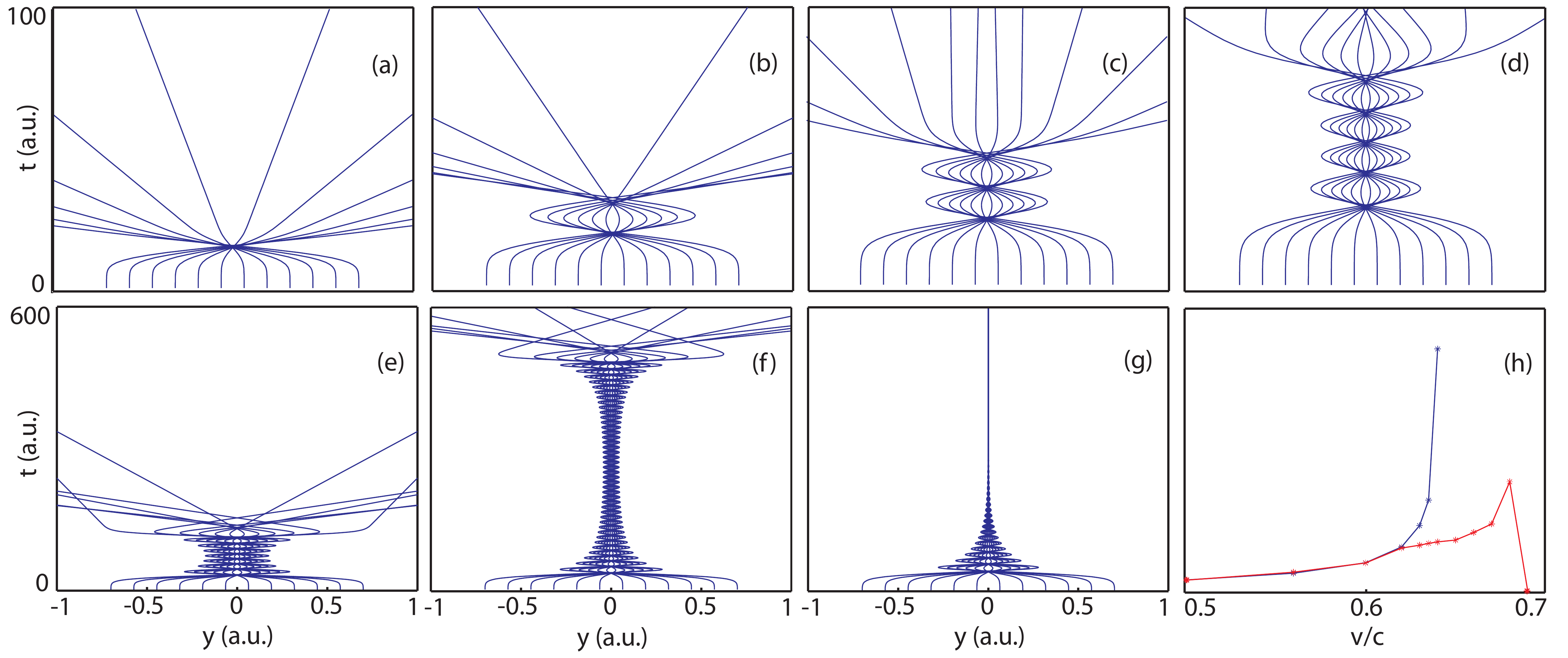}
\caption{Pictures (a) to (g) represent the trapping dynamics in the absence of dispersion for different RIP velocities: (a) $v=0.5$; (b) $v=0.56$; (c) $v=0.6$; (d) $v=0.62$; (e) $v=0.63$; (f) $v=0.64$; (g) $v=0.65$. Picture (h) represents the interaction time versus the velocity of the RIP in case of no dispersion (blue curve) and in case of dispersion (red curve).}\label{intrappolam}
\end{figure}
By increasing the velocity of the RIP from $v=0.5c$ up to $v=0.65c$ (i.e. approaching the trapping case), the rays spend more and more time inside the RIP. In Fig.~\ref{intrappolam}(a-g) we show this behaviour for different velocities, in the dispersion-less case. In  Fig.~\ref{intrappolam}(h), we have plotted the interaction time versus the velocity of the RIP in the dispersion (red curve) and dispersion-less (blue curve) case. Neglecting dispersion, the interaction time tends to infinity as the velocity of the RIP approaches the trapping case.  We can interpret this divergence as a (meta)stable state where the photon is trapped inside the pulse. An analytical estimate for this trapping time is given, for the axial geodesic by the formula $t_{trap}=\frac{2}{c} \int_0^r \frac{n d \xi}{1-n(\xi) v/c}$, where $r$ represents the longitudinal extension of the RIP.
In the dispersion-less case $t_{trap}$ diverges whereas, in the dispersive case, it is always finite.

\section{Horizon in optics}
\label{horiz}

\begin{figure}[ht]
\centering
\subfigure[Ergosphere and horizon.]{\label{opticalhorizon}\includegraphics[width=.4\linewidth]{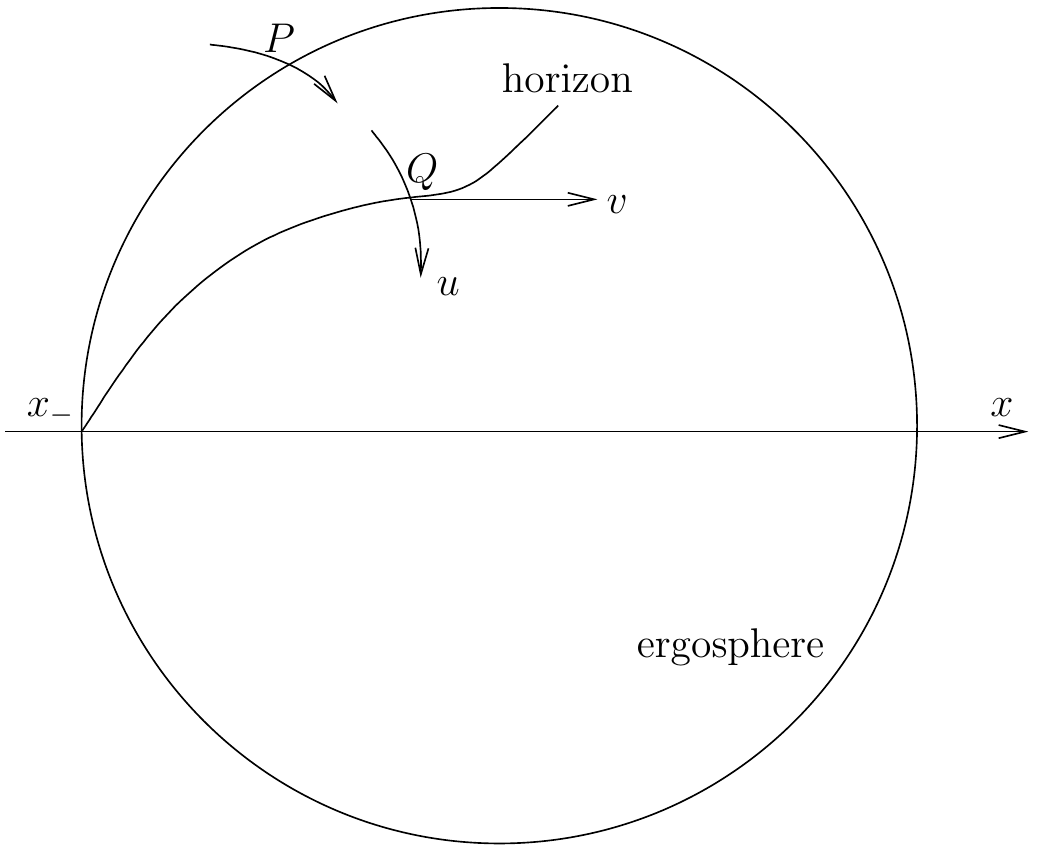}}
\qquad\qquad
\subfigure[Surface speed]{\label{opticalhorizon2}\includegraphics[width=.25\linewidth]{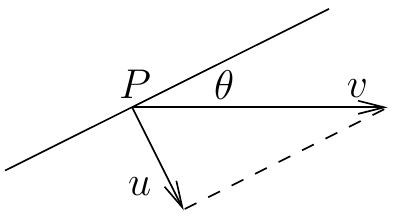}}
\caption{(a) Picture of the ergosphere and of the hypothetical horizon at a fixed
point of time in the laboratory frame. The small segments at $P$ and $Q$ represent
infinitesimal portions of null geodesics (light rays) passing, respectively, through
$P$ and $Q$ at that given point of time. (b) Normal velocity $u$ compared to the pulse
velocity, $v$.}
\end{figure}

In this section, we provide a definition for the horizons from the optical point of view, in the dispersion-less case. For the sake of definiteness we focus on the ``white hole'' horizon, but similar considerations hold true for the ``black hole'' horizon. We work in the laboratory frame, where the pulse is moving in the $x$ direction, with
velocity $v$. The situation is depicted in Fig.~\ref{opticalhorizon}.

Recall that the ergosphere is defined as the set of points where the velocity of light is the same as the velocity of the pulse, which means
$n(X)=c/v$ for all points $X$ on the ergosphere. In particular, it follows that a light ray reaching the ergosphere, for example in $P$, will
not enter the ball if it is moving in the $x$ direction. However, if it is directed along a NW-SE direction (for example toward the center)
then it will cross over the ergosphere thus entering the ball. This is because it ``sees'' the surface move slower in such a direction: indeed, note
that a plane forming an angle $\theta$ w.r.t. $x$ and moving with velocity $v$ in the $x$ direction, appears to move with velocity
$u=v\sin \theta<c/n(P)$ along its normal direction, see Fig.~\ref{opticalhorizon2}.

Therefore, only a single point on the ergosphere, namely the axial point $x_-$, behaves as a true white hole horizon. In order to determine whether a full event horizon extends beyond  $x_-$, we impose the condition $u=c/n(Q)$ on some point
internal to the ergosphere (where $n(Q)>n(P)=c/v$) which then gives
\begin{eqnarray}
v\sin \theta(Q) =c/n(Q). \label{hor-cond1}
\end{eqnarray}
Now, suppose that the horizon is described by a surface $\rho =h(x-vt)$, where $\rho$ is the radial coordinate. Then
\begin{eqnarray}
\tan \theta(Q)= \frac {dh}{dx} (Q). \label{hor-cond2}
\end{eqnarray}
Note that Eqs. (\ref{hor-cond1}) and (\ref{hor-cond2}) are indeed satisfied by the axial point $x_-$ (for which Eq. (\ref{hor-cond2}) gives $\theta=90^\circ$). A simple manipulation gives
\beq
\frac {dh}{dx}=\pm \frac c{\sqrt {v^2 n^2 -c^2}}.
\eeq
Finally, in the comoving frame, using coordinates $(x',\rho)$ the horizon equation is
\beq\label{eq-serg}
\frac {dh}{dx'}=\pm \frac{\sqrt{c^2-v^2}}{\sqrt {v^2 n^2 -c^2}}.
\eeq
The same equation can be obtained in the following way. It is a common strategy to
parametrise the horizon by means of a level-set function $F$ such that $F=0$
on the horizon; the lightlike condition for the event horizon is
$g^{\mu \nu} (\partial_\mu F)(\partial_\nu F)=0$.
In our case, there are two further requirements: axial symmetry and that the horizon is stationary (i.e. $\partial_{t' }F$=0). Then, in the RIP frame the condition which imposes the normal to the event horizon
to be lightlike is:
\beq\label{equa-light}
\gamma^2 \left(1-n^2 (x',\rho) \frac{v^2}{c^2} \right)
(\partial_{x'} F)^2 + (\partial_{\rho} F)^2 =0;
\eeq
in the static case, $\partial_{\rho} F=0$, and then, from (\ref{equa-light}) we deduce that the horizon consists of
two infinite extended planes, orthogonal to the $x$ axis, located at the solutions $x'_\pm$ of
$1-n^2 (x') \frac{v^2}{c^2}=0$,
as described in \cite{rip-static}.
\begin{figure}[t]
\centering
\includegraphics[width=0.4\textwidth]{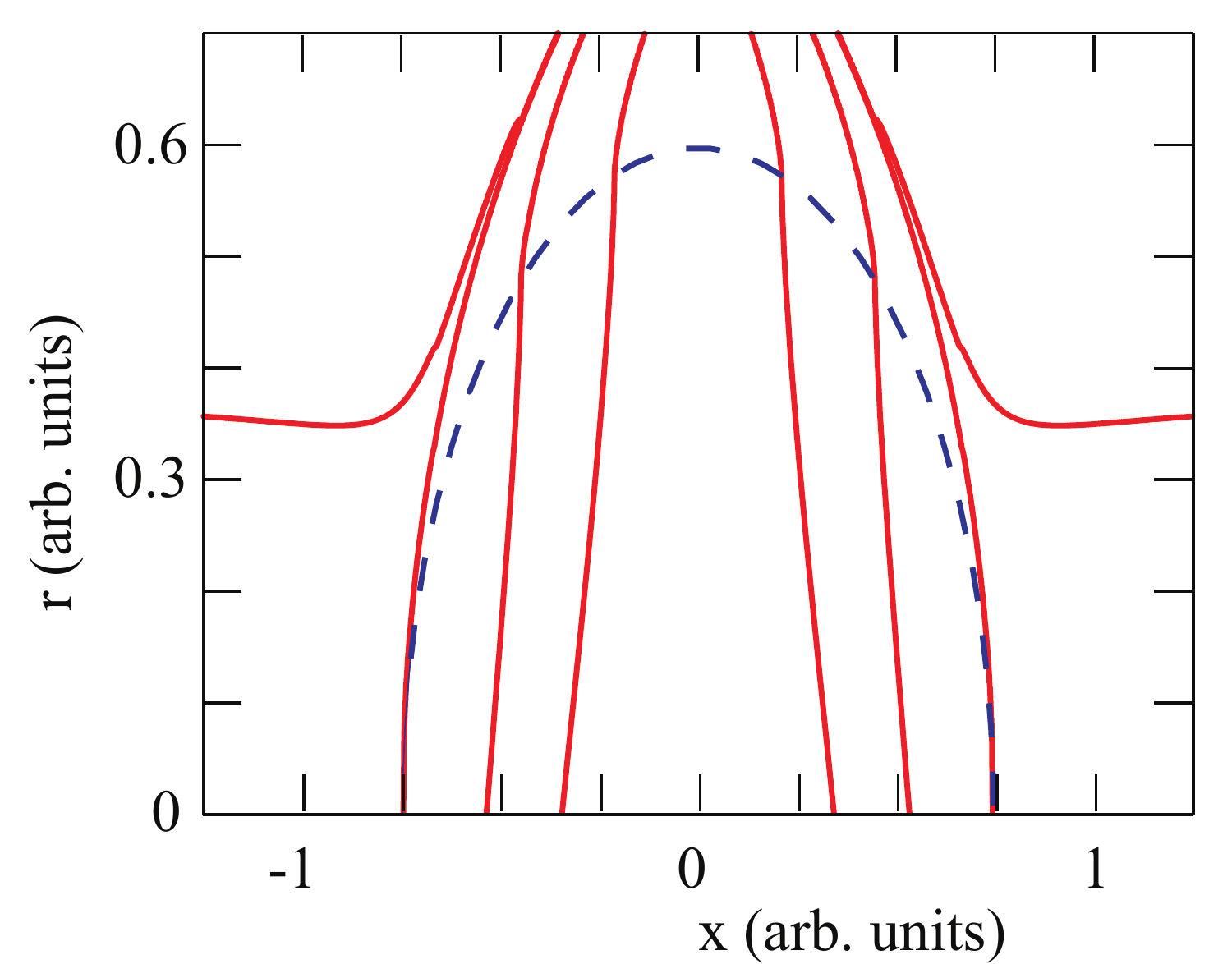}
\caption{Solid lines: real part of the solutions to Eq.~(\ref{eq-serg}). Dashed line: the ergosphere predicted by Eq.~(\ref{cond-ergo-eta}). The parameters are $n_0=1.45$, $\eta=0.1$, $v= c/1.52$.}\label{hor_eq_fig}
\end{figure}
In the nonstatic case, it is evident that the horizon has to be contained within the
ergosurface, because it is necessary that $1 - n^2 (x,\rho) v^2 /c^2 \leq 0$.
Moreover (at least locally, due to the implicit function theorem),
we assume that $F(x',\rho)=0 \Leftrightarrow \rho-h(x')=0$, i.e.
$F(x',\rho)=\rho-h(x')$. Then (\ref{equa-light}) becomes
$g^{11} (\frac{dh}{dx'})^2 + g^{22} =0$.
As a consequence, we obtain
$\frac{dh}{dx'} = \pm 1/\sqrt{-g^{11}}$,
which explicitly corresponds to (\ref{eq-serg}).

According to Eqs.~(\ref{hor-cond1}) and (\ref{hor-cond2}), as described above, the points $x_-$ and $x_+$ for $\rho=0$ belong to the event horizons, which then represent initial points for the
ordinary differential equation (\ref{eq-serg}). Allowed solutions should start from the aforementioned points and remain inside the ergosurface. According to numerical evaluations, carried out for $\eta = 1-10^{-4}$,
in the range of interest for our assumption that the RIP represents a
small perturbation with respect to the refractive index value in absence of the RIP,
the RIP exhibits a somewhat odd behaviour in the sense that the solution does not extend beyond $x_-$ and $x_+$.

Physically, a light ray passing through a point laying outside the ergoregion can travel in either direction (toward the RIP, or away from it). Such a ray is not ``trapped'', and therefore such a point cannot belong to an event horizon.

Figure~\ref{hor_eq_fig} shows the solutions to Eq.~(\ref{eq-serg}) for the case $\eta=0.1$. The dashed line shows the ergosurface predicted by Eq.~(\ref{cond-ergo-eta}) and the solid lines are solutions to  Eq.~(\ref{eq-serg}) for different initial conditions. These curves represent lightlike surfaces passing through points inside the ergosphere (in the RIP frame). The different null hypersurfaces which pass near $x_+$ (or $x_-$), cross the ergosphere at a point which is progressively closer to $x_+$ (or $x_-$). This is a numerical evidence of the fact that an hypothetical event horizon cannot extend beyond the points $x_+$ and $x_-$.
As can be seen, the solutions that pass through $x_+$ and $x_-$ are tangent to the ergosurface and always remain \emph{outside} the latter so that the horizon effectively reduces to a point.
We note that strictly speaking even though these points behave like points of a trapping horizon, as shown for example in Fig.~\ref{0p66}, it is of course not possible to define them as event horizons in the sense of a lightlike hypersurface.
One may wonder whether a ``trapping point'' is expected to still emit Hawking-like radiation and, if so,
with which features. In particular, can we rely on some arguments in order to
support the idea that there is a pair-creation process which, moreover, still shares common
features with the Hawking effect? We show in  a forthcoming work that the answer is affirmative
\cite{belg-black}. Points where the ergosurface becomes lightlike, beyond their properties to be
trapping points, which behave as standard horizons, share the further property to involve a production of
particles with thermal spectrum, again like in the case of standard (non-degenerate) horizons.

\section{Conclusions}

We have derived the effective metric associated with a RIP in a nonlinear Kerr medium. Such a metric  simulates
the effect on light rays of the presence of a dielectric perturbation in a sample of an isotropic $\chi^{(3)}$ dielectric
material. Then, we have investigated the possibility for an ergoregion to arise, and the behaviour of geodesics by means of numerical simulations, also taking into account the presence of dispersion. Finally, we have provided an equation for the event horizon that highlights a rather peculiar geometry, as it appears that the solutions may reduce to a single point. Numerical evaluation in the range of interest confirms the former conclusion. Still, the two special points $x_-$ and $x_+$ of the ergosurface, exhibit trapping properties like those of a standard trapping horizon
from a classical point of view.

\appendix

\section{A remark on the extraordinary metric}\label{appA}
We show that, under reasonable approximations, the extraordinary metric can be recast in the Gordon form.
We work in a reference frame at rest w.r.t. the medium (the laboratory frame).
Consider a background field $E$ which, in the laboratory frame, has polarization vector orthogonal to its direction of propagation $x$. We can choose Cartesian coordinates $(t,x,y,z)$ so that $l^\mu = (0,l_x = 0,l_y = 0,l_z=1)$.
The line element for the extraordinary rays takes the form
\beq
ds^2_{(-)} = \frac{c^2}{\tilde{n}^2} dt^2 - dx^2 - dy^2 -\left(1-\frac{\xi}{1+\xi}\right)dz^2 .
\eeq
Note that the factor in front of $dz^2$ cannot vanish. Moreover, $\xi \ll 1$ (see also \cite{rip-static}). Then, we can safely keep only the leading order term in the $z$ part of the metric so that, in so doing, the extraordinary metric reduces to
\begin{equation}
ds^2_{(-)} = \frac{c^2}{\tilde{n}^2} dt^2 - dx^2 - dy^2 - dz^2 .
\end{equation}
Under this approximation, the effective metric for extraordinary rays has again the same form as the Gordon one, with an effective refractive index
\begin{equation}
\tilde{n} = \sqrt{\mu_r\varepsilon_r(1+\xi)}= n_0 + \frac{3}{2}\frac{\sqrt {\mu_r}\chi^{(3)} E^2}{(1+\chi^{(1)})^{\frac 12}} = n_0 + 3 \delta n,
\end{equation}
to be compared to the refractive index for the ordinary rays, which is
\begin{equation}
n = \sqrt{\mu_r \varepsilon_r} =
n_0 + \frac{1}{2}\frac{\sqrt {\mu_r}\chi^{(3)} E^2}{(1+\chi^{(1)})^{\frac 12}} = n_0 + \delta n .
\end{equation}
Here, $n_0$ is the refractive index of the medium when we disregard nonlinear effects, i.e.
\begin{equation}
n_0 = \sqrt{\mu_r\varepsilon_r} = \sqrt{\mu_r\left(1+\chi^{(1)}\right)}.
\end{equation}
Then, the extraordinary rays ``feel'' a disturbance of the refractive index three times stronger than the ordinary ones.
The same conclusion can be also achieved in the context of Hawking radiation production by a RIP \cite{rip-static}. Indeed, it can be shown that the differential equation which defines the extraordinary field modes is well approximated by the same equation as for the ordinary ones, apart from an  effective RIP three times larger.
\newline

\bibliographystyle{unsrt}
\bibliography{stazionario}

\end{document}